\def\year{2015}
\documentclass[letterpaper]{article}
\usepackage{aaai}
\usepackage{times}
\usepackage{helvet}
\usepackage{courier}
\usepackage{graphicx}
\usepackage[linesnumbered,ruled,vlined]{algorithm2e}

\newtheorem{THEOREM}{Theorem}
\newenvironment{theorem}{\begin{THEOREM} }%
                        {\end{THEOREM}}
\newtheorem{LEMMA}[THEOREM]{Lemma}
                      {\end{LEMMA}}
\newtheorem{PROPOSITION}[THEOREM]{Proposition}
\newenvironment{prop}{\begin{PROPOSITION} }
                      {\end{PROPOSITION}}
\newtheorem{COROLLARY}[THEOREM]{Corollary}
                          {\end{COROLLARY}}
\newtheorem{EXAMPLE}{Example}
                            {\end{EXAMPLE}}

\newtheorem{DEFINITION}{Definition}
\newenvironment{definition}{\begin{DEFINITION} \rm }
                            {\end{DEFINITION}}

\newtheorem{PROBLEM}{Problem}

\newenvironment{proof}{{\it \noindent Proof:}}{\hfill\rule{2mm}{2mm} \vspace*{0.2cm}}

\newcommand{\ie}{\textit{i.e.}}
\newcommand{\eg}{\textit{e.g.}}
\newcommand{\lincom}{\textrm{LinCom}}
\newcommand{\davg}{d_{\emph{\tiny avg}}}
\newcommand{\dmax}{d_{\emph{\tiny max}}}

\begin{document}
%
\title{Exploiting Reduction Rules and Data Structures:\\ Local Search for Minimum Vertex Cover in Massive Graphs}
\author{
}
\maketitle
\begin{abstract}
\begin{quote}
The Minimum Vertex Cover (MinVC) problem is a well-known NP-hard problem. Recently there has been great interest in solving this problem on real-world massive graphs. For such graphs, local search is a promising approach to finding optimal or near-optimal solutions. In this paper we propose a local search algorithm that exploits reduction rules and data structures to solve the MinVC problem in such graphs. Experimental results on a wide range of real-word massive graphs show that our algorithm finds better covers than state-of-the-art local search algorithms for MinVC.
Also we present interesting results about the complexities of some well-known heuristics.
\end{quote}
\end{abstract}

\section{Introduction}

The Minimum Vertex Cover (MinVC) problem is a well-known NP-hard problem \cite{DBLP:conf/coco/Karp72} with many real-world applications \cite{Johnson:1996:CCS:548182}.
Given a simple undirected graph $G = (V, E)$ where $V$ is the vertex set and $E$ is the edge set. 
An edge $e$ is a set $\{u, v\}$ s.t. $u, v \in V$, and we say that $u$ and $v$ are endpoints of $e$.
A \emph{vertex cover} of a graph $G = (V, E)$ is a subset $V' \subseteq V$ s.t. for each $e \in E$, at least one of $e$'s endpoints is in $V'$.
The size of a vertex cover is the number of vertices in it.
The MinVC problem is to find a vertex cover of minimum size.

With growing interest in social networks, scientific computation networks and wireless sensor networks, etc., the MinVC problem has re-emerged even with greater significance and complexity, so solving this problem in massive graphs has become an active research agenda.
In this paper we are concerned in finding a vertex cover whose size is as small as possible.


It is hard to approximate MinVC within any factor smaller than 1.3606 \cite{DBLP:journals/ipl/DinurS04}.
During last decades there were many works in local search for MinVC like \cite{DBLP:conf/ki/RichterHG07,DBLP:journals/jair/CaiSLS13}.
Recently FastVC \cite{DBLP:conf/ijcai/Cai15} makes a breakthrough in massive graphs.
It makes a balance between the time efficiency and the guidance effectiveness of heuristics.
However, we realize that FastVC exploits very little about the structural information.
Also in order to achieve satisfactory time efficiency, it sacrifices the guidance effectiveness.

The aim of this work is to develop a local search MinVC solver to deal with massive graphs with strong structures.
The basic framework is this. 
Firstly, we exploit reduction rules to construct good starting vertex covers. 
Then we use local search to find better covers.
In both the construction stage and the local search stage, we exploit a novel data structure called \emph{alternative partitions} to pursue time efficiency, without sacrificing the quality of heuristics. 
Since we are now focusing on the impacts of the reduction rules and the data structures, we use naive local search strategies, so our solver may be too greedy.
For future works, we will exploit some strategies to diversify our local search.

Our solver constructs starting vertex covers by incorporating reduction rules.
In our experiments, our construction heuristic performs close to or even better than FastVC, on a large portion of the graphs.
Moreover, it outputs a cover typically within 10 seconds.
Hence, it provides a good starting point for later search.
Furthermore, for a small portion of the graphs, our heuristic guarantees that it has found an optimal cover, due to the power of reduction rules.
So far as we know, this is the first time reduction rules are applied in a local search MinVC solver, although they have been widely discussed in the community of theoretical computer science.

We also propose a brand new data structure to achieve time efficiency.
The main idea is to partition the vertices wrt. their scores, $\ie$, two vertices are in the same partition if they have the same score, otherwise they are in different partitions.
Thanks to this data structure, (1) as to the construction stage, the complexity of two important construction heuristics has been lowered, from $O(|V|^2)$ to $O(|V|+|E|)$; (2) as to the local search stage, the complexity of the best-picking heuristic has also been lowered, from $O(|C|)$ to $O(\davg)$ where $C$ is the set of vertices to be selected in local search, and $\davg$ is the average degree.
Later in this paper we will prove these results rigorously.
We applied these theoretical results in our solver, so we call our solver $\lincom$ (Linear-Complexity-Heuristic Solver).


We tested $\lincom$ and FastVC on the standard benchmark of massive graphs from the Network Data Repository\footnote{http://www.graphrepository.com./networks.php} \cite{rossi2015network}.
Our experiments show that among all the 12 classes of instances in this benchmark, $\lincom$ falls behind FastVC in only one class.
Moreover $\lincom$ finds smaller covers for a considerable portion of the graphs. 
This improvement is big, since it rarely happens in the literature \cite{DBLP:conf/ijcai/Cai15}.
\section{Preliminaries}
\subsection{Basic Notations}
If $e=\{u, v\}$ is an edge of $G$, we say that $u$ and $v$ are neighbors.
We define $N(v)$ as $\{u \in V | \{u, v\} \in E\}$.
The degree of a vertex $v$, denoted by $d(v)$, is defined as $|N(v)|$.
We use $\davg(G)$ and $\dmax(G)$ to denote the average degree and the maximum degree of graph $G$ respectively, suppressing $G$ if understood from the context.
An edge $e = \{u, v\}$ is covered by a vertex set $S$ if one of its endpoints is in $S$, $\ie$, $u \in S$ or $v \in S$ (or both).
Otherwise it is uncovered by $S$.

\subsection{Local Search for MinVC}

Most local search algorithms solve the MinVC problem by iteratively solving its decision version-given a positive integer $k$, searching for a $k$-sized vertex cover.
A general framework is Algorithm \ref{sls-mvc-framework}.
We denote the current candidate solution as $C$, which is a set of vertices selected for covering.

\begin{algorithm}
  \SetKwData{Left}{left}\SetKwData{This}{this}\SetKwData{Up}{up}
  \SetKwFunction{Union}{Union}\SetKwFunction{FindCompress}{FindCompress}
  \SetKwInOut{Input}{input}\SetKwInOut{Output}{output}
    $(C, optInfo) \leftarrow InitVC()$;\label{constr-stage}\\
    \While{not reach terminate condition}{\label{ls-stage-begin}
                \If{$C$ covers all edges}{\label{cover-found}
                    $C^* \leftarrow C$;\\
                    remove a vertex from $C$;\label{update-target-size}\\
                }
                    exchange a pair of vertices;\label{ls-stage-end}\\

    }
        \Return $C^*$
  \caption{A Local Search Framework for MinVC}\label{sls-mvc-framework}
\end{algorithm}

Algorithm \ref{sls-mvc-framework} consists of two stages: the construction stage (Line \ref{constr-stage}) and the local search stage (Line \ref{ls-stage-begin} to \ref{ls-stage-end}).
At the beginning, an initial vertex cover is constructed by the $InitVC$ procedure.
Throughout this paper, this initial cover is called the \emph{starting vertex cover}.
Besides $InitVC$ returns another parameter, $\ie$, $optInfo$ which takes the value \emph{optimal-guaranteed} or \emph{optimal-not-guaranteed} (See Algorithm \ref{init-vc}).

In the local search stage, each time a $k$-sized cover is found (Line \ref{cover-found}), the algorithm removes a vertex from $C$ (Line \ref{update-target-size}) and begins to search for a $(k-1)$-sized cover, until some termination condition is reached (Line \ref{ls-stage-begin}).
The move to a neighboring candidate solution consists of exchanging a pair of vertices (Line \ref{ls-stage-end}): a vertex $u \in C$ is removed from $C$, and a vertex $v \not\in C$ is added into $C$.
Such an exchanging procedure is also called a step by convention.
Thus the local search moves step by step in the search space to find a better vertex cover.
When the algorithm terminates, it outputs the smallest vertex cover that has been found.

For a vertex $v \in C$, the \emph{loss} of $v$, denoted as $loss(v)$, is defined as the number of covered edges that will become uncovered by removing $v$ from $C$.
For a vertex $v \not\in C$, the \emph{gain} of $v$, denoted as $gain(v)$, is defined as the number of uncovered edges that will become covered by adding $v$ into $C$.
Both loss and gain are \emph{scoring properties} of vertices.
In any step, a vertex $v$ has two possible states: inside $C$ and outside $C$, and we use $age(v)$ to denote the number of steps that have been performed since last time $v$'s state was changed.

\subsection{The Construction Stage}
Previous $InitVC$ procedures construct a starting vertex cover from an empty set $C$ mainly as below:
\begin{enumerate}
\item \textbf{Max-gain:} select a vertex $v$ with the maximum $gain$ and add $v$ into $C$, breaking ties randomly. Repeat this procedure until $C$ becomes a cover. \cite{PapadimitriousS82}
\item \textbf{Min-gain:} Select a vertex $v$ with the minimum positive $gain$ and add all $v$'s neighbors into $C$, breaking ties randomly. Repeat this procedure until $C$ becomes a cover. Redundant vertices (vertices whose $loss$ is 0) in $C$ are then removed. \cite{Ugrulu12ExcludeMinDegree,KettaniRBT13}
\item \textbf{Edge-greedy:} Select an uncovered edge $e$, add the endpoint with higher \emph{degree} into $C$. Repeat this procedure until $C$ becomes a cover. Redundant vertices in $C$ are then removed by a read-one procedure. \cite{DBLP:conf/ijcai/Cai15}
\end{enumerate}




\subsection{Reduction Rules for MinVC}
Our solver will incorporate the following reduction rules in the $InitVC$ procedure to handle vertices of small degrees.

\noindent\textbf{Degree-1 Rule:}
If $G$ contains a vertex $u$ s.t. $N(u)=\{v\}$, then there is a minimum vertex cover of $G$ that contains $v$.


The two rules below are from \cite{DBLP:journals/jal/ChenKJ01}.

\noindent\textbf{Degree-2 with Triangle Rule:}
If $G$ contains a vertex $v$ s.t. $N(v)=\{n_1, n_2\}$ and $\{n_1, n_2\} \in E$, then there is a minimum vertex cover of $G$ that contains both $n_1$ and $n_2$.



\noindent\textbf{Degree-2 with Quadrilateral Rule:}
If $G$ contains two vertices $u$ and $v$ s.t. $N(u)=N(v)=\{n_1, n_2\}$ and $\{n_1, n_2\}\not\in E$, then there is a minimum vertex cover of $G$ that contains both $n_1$ and $n_2$.


Since we are to develop a local search solver, we now rewrite them in the terminologies of local search.

\noindent\textbf{Degree-1 Rule:}
If $gain(v)=1$ and $u$ is a neighbor of $v$ s.t. $u \not\in C$, then put $u$ into the $C$.

\noindent\textbf{Degree-2 with Triangle Rule:}
If $gain(v)=2$, and $n_1,n_2$ are both $v$'s neighbors s.t. $n_1, n_2\not\in C$ and $\{n_1, n_2\}\in E$, then put both $n_1$ and $n_2$ into the $C$.

\noindent\textbf{Degree-2 with Quadrilateral Rule:}
If $gain(u)$ = $gain(v)$ = $2$, and both $n_1,n_2$ are neighbors shared by $u,v$ s.t. $n_1, n_2 \not\in C$ and $\{n_1, n_2\}\not\in E$, then put both $n_1$ and $n_2$ into the $C$.

\section{Incorporating Reduction Rules}
We incorporate reduction rules in order to: (1) construct smaller starting vertex covers; (2) help confirm optimality.
\subsection{Constructing A Vertex Cover with Reductions}
Like \cite{DBLP:conf/ijcai/Cai15}, our $InitVC$ procedure also consists of an extending phase (Lines \ref{extend-phase-begin} to \ref{max-gain-pick}) and a shrinking phase (Line \ref{shrink}).
Notice that if we construct a cover by only using reduction rules, then it must be optimal.
So we employ a predicate \emph{max\_gain\_used} s.t. \emph{max\_gain\_used} = \emph{true} if Line \ref{max-gain-select} has been executed, and \emph{max\_gain\_used} = \emph{false} otherwise.

\begin{algorithm}
  \SetKwData{Left}{left}\SetKwData{This}{this}\SetKwData{Up}{up}
  \SetKwFunction{Union}{Union}\SetKwFunction{FindCompress}{FindCompress}
  \SetKwInOut{Input}{input}\SetKwInOut{Output}{output}

  \Input{A graph $G = (V, E)$}
  \Output{A cover $C$ and whether-optimal-guaranteed}
  \BlankLine
    $C \leftarrow \emptyset$;\label{init-empty-C}\\
    \emph{max\_gain\_used} $\leftarrow$ \emph{false};\label{max-gain-used-init}\\
    \While{there exist uncovered edges}
    {\label{extend-phase-begin}
        Repeatedly apply the \emph{Degree-2 with Triangle Rule} until it is not applicable;\label{tri-rule}\\
        Repeatedly apply the \emph{Degree-2 with Quadrilateral Rule} until it is not applicable;\label{qua-rule}\\
        Repeatedly apply the \emph{Degree-1 Rule} until it is not applicable;\label{degree-1-rule}\\
        \lIf{any rule above is applicable}{continue}\label{check-inference-rule-applicable}
        \lIf{all edges are covered}{break}\label{cover-formed}
        \emph{max\_gain\_used} $\leftarrow$ \emph{true};\label{max-gain-used-set}\\
                pick a vertex $v$ with the maximum $gain$ (ties are broken randomly), put it into $C$;\label{max-gain-select}\label{max-gain-pick}
    }
    $C\leftarrow$ eliminateRedundantVertices($C$);\label{shrink}\\
    \eIf{max\_gain\_used = true}{\label{max-gain-used-decide}
        \Return ($C$, \emph{optimal-not-guaranteed});
        }
        {
        \Return ($C$, \emph{optimal-guaranteed});
        }
  \caption{InitVC}\label{init-vc}
\end{algorithm}

In Line \ref{init-empty-C}, we initialize $C$ to be an empty set.
Then we extend $C$ to be a vertex cover of $G$, by iteratively adding a vertex into $C$.
Lines \ref{tri-rule} to \ref{degree-1-rule} apply reduction rules to put vertices into $C$.
Line \ref{check-inference-rule-applicable} ensures that no reduction rules are applicable before making use of the max-gain heuristic.
After the extending phase (Lines \ref{extend-phase-begin} to \ref{max-gain-pick}), Line \ref{shrink} removes the redundant vertices from $C$ just as what \cite{DBLP:conf/ijcai/Cai15} did.

\subsection{Fixing Vertices in the Starting Vertex Cover}
When Algorithm \ref{init-vc} constructs a starting vertex cover, we realize that some of the vertices are put into $C$ based on pure reductions.
That is, they were put into $C$ when \emph{max\_gain\_used = false}.
Hence, there exist a minimum vertex cover which contains all of such vertices, and we call them \emph{inferred vertices}.
In local search we can fix the inferred vertices in $C$ s.t. they are never allowed to be removed from $C$.
It seems that such a procedure are able to reduce the search space and speed up the search.

So we employ an array $fixed$, whose element is an indicator for a vertex.
During the execution of Algorithm \ref{init-vc}, we maintain the $fixed$ array as below:
\begin{enumerate}
\item Rule 1: Before the extending phase, for each vertex $v$, $fixed[v]$ is set to \emph{false}.
    \item Rule 2: When putting a vertex into $C$, we check whether \emph{max\_gain\_used = false}. If so, $fixed[v]$ is set to \emph{true}.
\end{enumerate}
Thus when Algorithm \ref{init-vc} is completed, $fixed[v]$\emph{ = true} if $v$ is an inferred vertex, and $fixed[v]$\emph{ = false} otherwise.
So later when we are doing local search, we can forbid $u$ from being removed from $C$ if $fixed[u]$ = \emph{true}, as is shown in Line \ref{remove-free-v} and \ref{choose-free-v} in Algorithm \ref{lincom}.

\section{A Local Search MinVC Solver}
\begin{algorithm}
  \SetKwData{Left}{left}\SetKwData{This}{this}\SetKwData{Up}{up}
  \SetKwFunction{Union}{Union}\SetKwFunction{FindCompress}{FindCompress}
  \SetKwInOut{Input}{input}\SetKwInOut{Output}{output}
  \Input{A graph $G = (V, E)$, the \emph{cutoff} time}
  \Output{A vertex cover of $G$}
  \BlankLine
    $(C, optInfo) \leftarrow InitVC()$;\label{}\\
    \lIf{$optInfo = $ optimal-guaranteed}{\Return $C$}
    \While{elapsed time $<$ cutoff}{\label{}
                \If{$C$ covers all edges}{\label{}
                    $C^* \leftarrow C$;\\
                    remove a vertex $u$ s.t. $fixed[u]=$ \emph{false} with minimum $loss$ from $C$, breaking ties randomly;\label{remove-free-v}\\
                }
                    remove a vertex $u \in C$ s.t. $fixed[u]=$ \emph{false} with the minimum $loss$, breaking ties randomly;\label{choose-free-v}\\
                    $e \leftarrow$ a random uncovered edge;\\
                    add the endpoint of $e$ with the greater $gain$, breaking ties in favor of the older one;\\

    }
        \Return $C^*$;
  \caption{$\lincom$($G$, \emph{cutoff})}\label{lincom}
\end{algorithm}


Our solver $\lincom$ is outlined in Algorithm \ref{lincom}.
At first a vertex cover is constructed.
If the returned cover is guaranteed to be optimal, the algorithm will immediately return.

Then at each step, the algorithm first chooses a vertex $u\in C$ s.t. $u$ is not an \emph{inferred vertex} ($\ie$, $fixed[u]$ = \emph{false}) with the minimum $loss$, breaking ties randomly.
Then the algorithm picks a random uncovered edge $e$, chooses one of $e$'s endpoints with the greater $gain$ and adds it, breaking ties in favor of the older one.

\section{Data Structures}



In order to lower the complexities, we exploited an efficient data structure named \emph{alternative partitions} (See Figure \ref{add-v-1}).
\subsection{Alternative Partitions}

We use \emph{loss}-$k$ (resp. \emph{gain}-$k$) partition to denote the partition that contains vertices in $C$ (resp. outside $C$) whose loss (resp. gain) is $k$ (Figure \ref{add-v-1}).
All the \emph{loss}-$k$ partitions are shown as dark regions, and all the \emph{gain}-$k$ partitions are shown as light ones.
Since the dark and the light regions are distributed alternatively, we call them \emph{alternative partitions}.
Obviously we have
\begin{prop}
\begin{enumerate}\label{gain-loss-bound-prop}
\item $0 \leq gain(v) \leq d(v) \leq |V|$ where $v \not\in C$.
\item $0 \leq loss(v) \leq d(v) \leq |V|$ where $v \in C$.
\end{enumerate}
\end{prop}
Then we use Algorithm \ref{best-pick} to find those vertices in $C$ with the minimum loss.
\begin{algorithm}
  \SetKwData{Left}{left}\SetKwData{This}{this}\SetKwData{Up}{up}
  \SetKwFunction{Union}{Union}\SetKwFunction{FindCompress}{FindCompress}
  \SetKwInOut{Input}{input}\SetKwInOut{Output}{output}

  \Input{A sequence of alternative partitions}
  \Output{A random vertex $v \in C$ with minimum loss}
  \BlankLine
    $k \leftarrow 0$;\\
    \lWhile{the loss-$k$ partition is empty}{
            $k \leftarrow k + 1$
        }
        \Return a random vertex in the \emph{loss}-$k$ partition;
  \caption{randomMinLossVertex}\label{best-pick}
\end{algorithm}

In this algorithm we first check whether there are any vertices whose loss is 0.
If so, we randomly return one of them.
Otherwise, we go on to check whether there are any vertices whose loss is 1, 2, $\ldots$ until we find a non-empty partition.
Then we randomly return one in that partition.
So we have,
\begin{prop}\label{locate-min-loss-prop}
The complexity of Algorithm \ref{best-pick} is $O(\dmax)$.
\end{prop}

Similarly we have
\begin{prop}\label{locate-min-gain-prop}
The complexity of finding the partition with the maximum/minimum gain is $O(\dmax)$.
\end{prop}

\subsection{Implementations}
Given a graph $G = (V, E)$ and a candidate solution $C$, we implement the alternative partitions on an array where each position holds a vertex (See Figure \ref{add-v-1}).
Besides, we maintain two additional arrays of pointers, each of which points to the beginning of a specific partition.
Imagine the array as a book of vertices and the pointer arrays as the indexes of the book.
\subsubsection{Initializing the Partitions}
At first when $C$ is empty, there are no dark regions in our data structure, so initializing the partitions is equivalent to sorting the vertices into a monotonic nondecreasing order, based on their gain.
Notice that at this time, the gain of any vertex is equal to its degree, so we now need to sort vertices by degrees.
By Proposition \ref{gain-loss-bound-prop}, 
this satisfies the assumption of counting sort which runs in linear time \cite{DBLP:books/daglib/introToAlgo}.
Thus we have,
\begin{prop}\label{init-partition-prop}
Initializating the partitions is $O(|V|)$.
\end{prop}
\subsubsection{Maintaining the Partitions}
After initializations, there are two cases in which a particular vertex, say $v$, has to be moved from one partition to another: (1) adding (resp. removing) $v$ into (resp. from) $C$; (2) increasing/decreasing $gain(v)$/$loss(v)$
by 1.
Thus the core operation is to move a vertex $v$ to an adjacent partition.

\begin{center}
\begin{figure}[h]
\includegraphics[totalheight=1.0in]{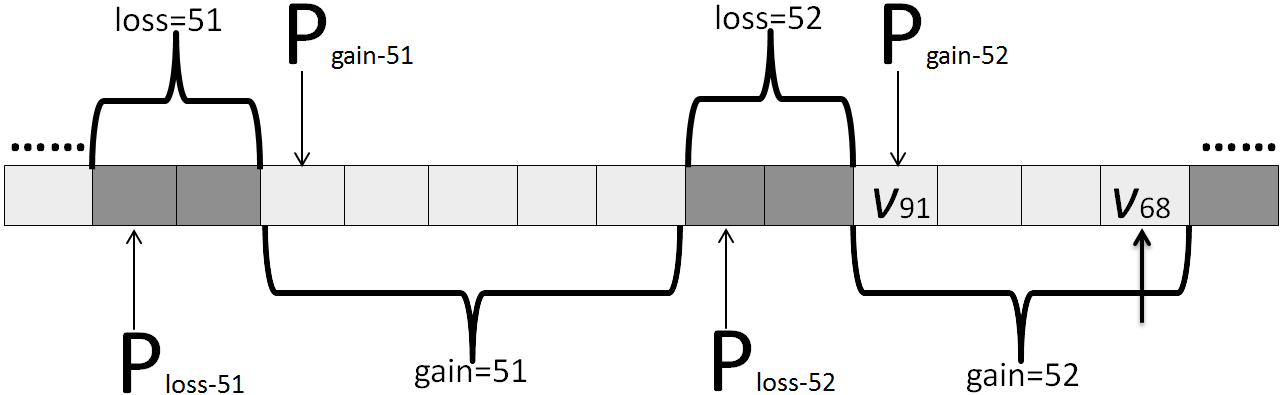}
\caption{Adding $v_{68}$ into $C$ (a)}
\label{add-v-1}
\end{figure}
\end{center}

\begin{center}
\begin{figure}[h]
\includegraphics[totalheight=1.0in]{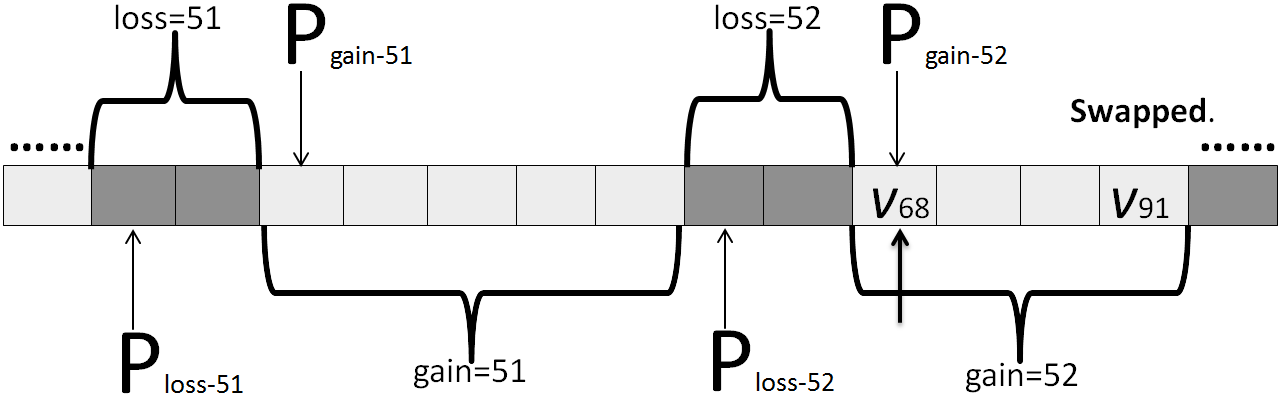}
\caption{Adding $v_{68}$ into $C$ (b)}
\label{add-v-2}
\end{figure}
\end{center}

\begin{center}
\begin{figure}[h]
\includegraphics[totalheight=1.01in]{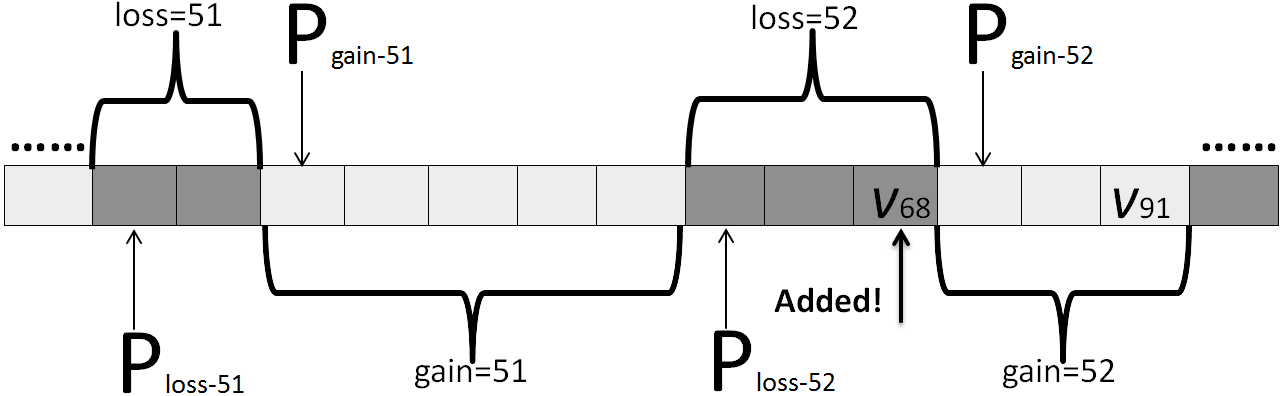}
\caption{Adding $v_{68}$ into $C$ (c)}
\label{add-v-3}
\end{figure}
\end{center}

Now we show how to do this with an example (See Figure \ref{add-v-1} to \ref{add-v-3}).
In this example, we are to add $v_{68}$ into $C$.
Initially $v_{91}$ and $v_{68}$ are in the \emph{gain}-52 partition and thus their \emph{gain} is 52 (Figure \ref{add-v-1}).
Notice that after being added, $v_{68}$'s loss will become 52, $\ie$, it should be in the \emph{loss}-52 partition.
Thus the operation is performed like this: (1) $v_{68}$ is swapped with $v_{91}$ (Figure \ref{add-v-2}); (2) \textsf{P}$_{\textsf{\tiny gain-52}}$ is moved (Figure \ref{add-v-3}).

We define placeVertexIntoC($v$) as the procedure that moves $v$ from certain \emph{gain}-$k$ partition to the respective \emph{loss}-$k$ partition, puts it into $C$ and updates its score.
And we define gainMinusMinus($v$) as the procedure that moves $v$ from certain \emph{gain}-$k$ partition to the respective \emph{gain}-$(k-1)$ partition and updates its score.
Analogously we define placeVertexOutfromC($v$), lossMinusMinus($v$), gainPlusPlus($v$), and lossPlusPlus($v$).
Then we have
\begin{prop}\label{constant-basic-op}
All the procedures are of $O(1)$ complexities.
\end{prop}

\section{Complexity Analysis}
In this section, we evaluate the complexities of the best-picking and the vertex cover construction heuristics.
\subsection{Complexity of The Best-picking Heuristic}
Along with adding/removing a vertex $v$, we have to move this vertex and all its neighbors to other partitions.
Thus by Proposition \ref{constant-basic-op}, maintaining the partitions will take $O(1)$ time plus an amount of time proportional to $d(v)$.
Thus,
\begin{prop}\label{maintain-bucket-prop}
When a vertex is added/removed, the complexity of maintaining the partitions at each step is $O(\dmax)$.
\end{prop}

By Proposition \ref{locate-min-loss-prop} and \ref{maintain-bucket-prop}, we have
\begin{prop}
The best-picking heuristic in Algorithm \ref{lincom} can be done in $O(\dmax)$ complexity.
\end{prop}

In the local search stage, by Proposition \ref{gain-loss-bound-prop}, we have
\begin{theorem}
Suppose that each vertex has equal probability to be added or removed, then the average complexity of the best-picking heuristic in Algorithm \ref{lincom} is $O(\davg)$.
\end{theorem}

It is nice because \cite{DBLP:conf/ijcai/Cai15} stated that the best-picking heuristic was of $O(|C|)$ complexity.
Since most real-world graphs are sparse \cite{Barabasi99emergenceScaling,DBLP:conf/soda/EubankKMSW04,chung2006complex}, we have $\davg << |C|$.

\subsection{Complexity of The Max-gain/Min-gain Heuristics}
\begin{algorithm}
  \SetKwData{Left}{left}\SetKwData{This}{this}\SetKwData{Up}{up}
  \SetKwFunction{Union}{Union}\SetKwFunction{FindCompress}{FindCompress}
  \SetKwInOut{Input}{input}\SetKwInOut{Output}{output}

  \Input{A graph $G = (V, E)$}
  \Output{A cover $C$ and whether-optimal-guaranteed}
  \BlankLine
    $C \leftarrow \emptyset$; $|UE| \leftarrow |E|$;
    initialize the partitions;\label{init-partition-exec}\\
    \While{$|UE| > 0$}{
        $k \leftarrow 1$;\\
        \lWhile{the gain-$k$ partition is empty}{
            $k \leftarrow k + 1$
        }\label{find-min-gain-partition}
        $min\_g\_v \leftarrow $ a random vertex in the \emph{gain}-$k$ partition;\label{rand-min-gain-v}\\
        \ForEach{$v \in N(min\_g\_v)$}{
        	\lIf{$v \in C$}{continue}\label{test_nb_to_put}
            placeVertexIntoC($v$); $|UE| \leftarrow |UE| - gain(v)$;\label{placeIn}\\
            \ForEach{$n \in N(v)$}{
                \lIf{$n \in C$}{\label{test_v_in_c}
                    lossMinusMinus($n$)
                }\label{loss-minus-minus}
                \lElse{
                    gainMinusMinus($n$)
                }\label{gain-minus-minus}
            }
        }
    }
        \Return ($C$, \emph{optimal-not-guaranteed});
  \caption{minGainConstructVC}\label{min-gain-init}
\end{algorithm}

\cite{DBLP:conf/ijcai/Cai15} formally proved that the max-gain heuristic had a worst-case complexity of $O(|V|^2)$.
Moreover, both \cite{Ugrulu12ExcludeMinDegree} and \cite{KettaniRBT13} proved rigorously that the worst-case complexity of the min-gain heuristic was $O(|V|^2)$.
Yet with the alternative partitions, we have
\begin{theorem}
The min-gain/max-gain heuristic constructs a vertex cover in $O(|V|+|C|+|E|)$ complexity, where $C$ is the starting vertex cover.
\end{theorem}

\vspace{2mm}
\begin{proof}
We use $UE$ to denote the set of uncovered edges.
\begin{enumerate}
\item We prove the case for min-gain by Algorithm \ref{min-gain-init}.
By Proposition \ref{init-partition-prop}, Line \ref{init-partition-exec} has a complexity of $O(|V|)$.

In any cycle of the outer while-loop, if the condition in Line \ref{find-min-gain-partition} is tested for $t$ times, then $gain(min\_g\_v)=t$, and thus $t$ neighbors of $min\_g\_v$ will be put into $C$.
That is, in any cycle, the number of tests done in Line \ref{find-min-gain-partition} is equal to the number of vertices that will be put. 
So that condition will be tested for exactly $|C|$ times during the algorithm.

Given $min\_gain\_v$ in Line \ref{rand-min-gain-v}, the algorithm tests each of its neighbors whether they are in $C$ in Line \ref{test_nb_to_put}.
Considering the case that we have to test every neighbor of every vertex, the total number of tests done is $2|E|$.
Thus the condition in Line \ref{test_nb_to_put} will be tested for at most $2|E|$ times.

After putting a vertex into $C$ in Line \ref{placeIn}, we have to update the information about its neighbors (Line \ref{loss-minus-minus}-\ref{gain-minus-minus}). Again considering the extreme case above, the total number of updates (gainMinusMinus or lossMinusMinus) will be at most $2|E|$. 
By Proposition \ref{constant-basic-op}, the time spent in Line \ref{loss-minus-minus}-\ref{gain-minus-minus} during the algorithm is $O(|E|)$.
To conclude, the overall complexity is $O(|V|+|C|+|E|)$.

\item We prove the case for max-gain by Algorithm \ref{max-gain-init}.

In Line \ref{init-max-gain-k}, we initialize $k$ to be $\dmax$ which is equal to the maximum gain at this time.
Notice that the value of the maximum gain never increases in the construction stage.
So during the execution, whenever we find that there are no vertices whose gain is $g$, we go on to check whether there  are any vertices whose gain is $g-1$.
Thus, during the execution, $k$ is always the value of the maximum gain.

When the condition in Line \ref{find-max-gain-partition} is tested, there are two cases: (1) if succeeds, then $k$ is decreased by 1; (2) if fails, then one vertex is put into $C$.
So the number of tests done in Line \ref{find-max-gain-partition} is exactly $|C|+\dmax \leq |C|+|V|$.
Similarly, the overall complexity is $O(|V|+|C|+|E|)$.
\end{enumerate}
\end{proof}

\begin{algorithm}
  \SetKwData{Left}{left}\SetKwData{This}{this}\SetKwData{Up}{up}
  \SetKwFunction{Union}{Union}\SetKwFunction{FindCompress}{FindCompress}
  \SetKwInOut{Input}{input}\SetKwInOut{Output}{output}

  \Input{A graph $G = (V, E)$}
  \Output{A cover $C$, whether-optimal-guaranteed}
  \BlankLine

    $C \leftarrow \emptyset$; $|UE| \leftarrow |E|$; initialize the partitions;\\
    $k \leftarrow \dmax$;\label{init-max-gain-k}\\
    \While{$|UE| > 0$}{

                \lWhile{the gain-$k$ partition is empty}{$k \leftarrow k - 1$}\label{find-max-gain-partition}
                 $v \leftarrow $a random vertex in the \emph{gain}-$k$ partition;\\
                 placeVertexIntoC($v$); $|UE| \leftarrow |UE| - gain(v)$;\label{placeIn-max-gain}\\
            \ForEach{$n \in N(v)$}{
                \lIf{$n \in C$}{\label{test_v_in_c}
                    lossMinusMinus($n$)
                }\label{loss-minus-minus-max-gain}
                \lElse{
                    gainMinusMinus($n$)
                }\label{gain-minus-minus-max-gain}
            }

    }
        \Return ($C$, \emph{optimal-not-guaranteed});
  \caption{maxGainConstructVC}\label{max-gain-init}
\end{algorithm}

Besides, we compared Algorithm \ref{max-gain-init} with the traditional one \cite{DBLP:journals/jair/CaiSLS13} through experiments.
Moreover, as to the min-gain heuristic, we program it ourselves in two ways: Algorithm \ref{min-gain-init} and the previous way.
It shows that our methods are faster than the traditional ones by orders of magnitude on large instances.
So our experimental results were completely consistent with the theoretical expectations.
So far we have not derived the complexity of Algorithm \ref{init-vc} yet, but we believe that it is also linear, because our $InitVC$ procedure outputs a vertex cover typically within 10 seconds.  

Because the max-gain heuristic was proposed about three decades ago \cite{PapadimitriousS82}, and \cite{DBLP:conf/ijcai/Cai15} still proved the $O(|V|^2)$ complexity, our result is surprising.
Note that partitioning is a general method and can also be applied to solve huge instances for other problems.

\section{Experimental Evaluation}

In this section, we carry out extensive experiments to evaluate $\lincom$ on massive graphs, compared against the state-of-the-art local search MinVC algorithm FastVC.
To show the individual impacts, we also present the performances of our $InitVC$ procedure (named as InitVC in the tables).

\subsection{Benchmarks}

We downloaded all 139 instances\footnote{http://lcs.ios.ac.cn/\~caisw/Resource/realworld\%20graphs.tar.gz}.
They were originally online,\footnote{http://www.graphrepository.com./networks.php} and then transformed to DIMACS graph format.
But we excluded three extremely large ones, since they are out of memory for all the algorithms here.
Thus we tested all the solvers on the remaining 136 instances.
Some of them have recently been used in testing parallel algorithms for Maximum Clique and Coloring problems \cite{DBLP:journals/snam/RossiA14,DBLP:conf/www/RossiGGP14}.


\subsection{Experiment Setup}

All the solvers were compiled by g++ 4.6.3 with the '-O3' option.
For FastVC\footnote{http://lcs.ios.ac.cn/~caisw/Code/FastVC.zip}, we adopt the parameter setting reported in \cite{DBLP:conf/ijcai/Cai15}.
The experiments were conducted on a cluster equipped with a number of Intel(R) Xeon(R) CPUs
X5650 @2.67GHz with 8GB RAM, running Red Hat Santiago OS.

All the algorithms are executed on each instance with a time limit of 1000 seconds, with seeds from 1 to 100.
For each algorithm on each instance, we report the minimum size ("$C_{min}$") and averaged size ("$C_{avg}$") of vertex covers found by the algorithm.
To make the comparisons clearer, we also report the difference ("$\Delta$") between the minimum size of vertex cover found by FastVC and that found by $\lincom$.
A positive $\Delta$ means that $\lincom$ finds a smaller vertex cover, while a negative $\Delta$ means that FastVC finds a smaller vertex cover.
The numbers of vertices of these graphs lie between $1\times 10^3$ to $4\times 10^6$.
We omit them and readers may refer to \cite{DBLP:conf/ijcai/Cai15} or the download website.
\subsection{Experimental Results}
We show the main experimental results in Tables \ref{tab:1} and \ref{tab:2}.
For the sake of space, we do not report the results on graphs with less than 1000 vertices.
Furthermore, we do not report the results on graphs where $\lincom$ and FastVC precisely return both the same minimum size and average size.
\begin{table} \tiny
\setlength{\tabcolsep}{2pt}
\renewcommand{\arraystretch}{1}
    \caption{Experimental results on collaboration networks, facebook networks, interaction networks, infrastructure networks, recommend networks and retweet networks}
    \begin{tabular}{| l | l | l | l | l |}
    \hline
    Graph & FastVC  & InitVC & $\lincom$ & $\Delta $ \\
     & $C_{min}(C_{avg})$ & $C_{min}(C_{avg})$ & $C_{min}(C_{avg})$& \\\hline \hline
    ca-AstroPh & 11483 (11483) &11483 (11483.36)  & 11483 (11483.01)& 0 \\
    ca-citeseer & 129193 (129193) &129193 (129193.82)  & 129193 (129193.36)& 0 \\
    ca-coauthors-dblp & 472179 (472179) & 472234 (472242.19) & 472179 (472179.02)& 0 \\
    ca-CondMat & 12480 (12480) & 12481 (12481.25) & 12480 (12480.06)& 0 \\
    ca-dblp-2010 & 121969 (121969) & 121970 (121971.02) & 121969 (121969.64)& 0 \\
    ca-dblp-2012 & 164949 (164949) &164949 (164950.88)  & 164949 (164950.35)& 0 \\
    ca-hollywood-2009 & 864052 (864052) & 864052 (864053.9) & 864052 (864052.01)& 0 \\
    ca-MathSciNet & 139951 (139951) & 139951 (139952.45) & 139951 (139952.23)& 0 \\\hline
    socfb-A-anon & 375231 (375232.94) & 375230 (375230.82) & 375230 (375230.82)& 1 \\
    socfb-B-anon & 303048 (303048.93) & 303048 (303048) & 303048 (303048)& 0 \\
    socfb-Berkeley13 & 17209 (17212.18) & 17280 (17290.32 ) & 17210 (17215.93)& -1 \\
    socfb-CMU  & 4986 (4986.72) & 5002 (5007.41) & 4986 (4987.24)& 0 \\
    socfb-Duke14 & 7683 (7683.05) & 7707 (7712.34) & 7683 (7684.98)& 0 \\
    socfb-Indiana & 23313 (23317.19) & 23426 (23439.12) & 23319 (23323.79)& -6 \\
    socfb-MIT & 4657 (4657) & 4663 (4669.13) & 4657 (4657.56)& 0 \\
    socfb-OR & 36547 (36549.44 ) & 36586 (36594.26) & 36548 (36549.50)& -1 \\
    socfb-Penn94 & 31161 (31164.95) & 31299 (31313.34) & 31165 (31170.78)& -4 \\
    socfb-Stanford3 & 8517 (8517.89) & 8534 (8540.01) & 8518 (8518.35)& -1 \\
    socfb-Texas84 & 28166 (28171.54) & 28306 (28317.76) & 28169 (28178.98)& -3 \\
    socfb-UCLA & 15222 (15224.41) & 15279 (15294.25) & 15224 (15228.85)& -2 \\
    socfb-UConn & 13230 (13231.60) & 13287 (13300.16) & 13232 (13235.99)& -2 \\
    socfb-UCSB37 & 11261 (11262.88) & 11310 (11316.65) & 11262 (11265.54)& -1 \\
    socfb-UF & 27305 (27309.04) & 27440 (27453.23) & 27310 (27316.25)& -5 \\
    socfb-UIllinois & 24090 (24093.97) & 24209 (24222.07) & 24095 (24101.18)& -5 \\
    socfb-Wisconsin87 & 18383 (18385.46) & 18468 (18483.70) & 18384 (18390.13)& -1 \\ \hline
    ia-enron-large & 12781 (12781) & 12781 (12781.2) & 12781 (12781.2)& 0 \\ \hline
    inf-power & 2203 (2203) & 2203 (2203.01) & 2203 (2203.01)& 0 \\
    inf-roadNet-CA & 1001254 (1001325.29) & 1007098 (1007362.34) & 1001058 (1001139.61)& 196 \\
    inf-roadNet-PA & 555203 (555248.74) & 558206 (558343.72) & 555035 (555107.22)& 168 \\ \hline
    rec-amazon & 47606(47606.01) & 47605 (47611.64) & 47605 (47605.62)& 1 \\ \hline
    rt-retweet-crawl & 81044 (81047.81) & 81040 (81040) & 81040 (81040)& 4 \\ \hline

    \hline
    \end{tabular}
    \label{tab:1}
\end{table}

\begin{table} \tiny
\setlength{\tabcolsep}{2pt}
\renewcommand{\arraystretch}{1}
	\caption{Experimental results on scientific computation networks, social networks, technological networks, temporal reachability networks and web link networks}
    \begin{tabular}{| l | l | l | l | l |}
    \hline
    Graph & FastVC  & InitVC & $\lincom$ & $\Delta $ \\
     & $C_{min}(C_{avg})$ & $C_{min}(C_{avg})$ & $C_{min}(C_{avg})$& \\\hline \hline
    sc-ldoor & 856754 (856757.36) & 858142 (858173.08) & 856755 (856757.18)& -1 \\
    sc-msdoor & 381558 (381559.23) & 382102 (382120.66) & 381559 (381559.86)& -1 \\
    sc-nasasrb & 51242 (51247.27) & 51575 (51605.64) & 51243 (51249.23)& -1 \\
    sc-pkustk11 & 83911 (83912.97) & 84124 (84146.02) & 83911 (83913.52)& 0 \\
    sc-pkustk13 & 89217 (89220.46) & 89625 (89652.49) & 89219 (89222.95)& -2 \\
    sc-pwtk & 207711 (207720.22) & 208713 (208760.96) & 207698 (207711.11)& 13 \\
    sc-shipsec1 & 117305 (117338.65) & 118727 (118788.57) & 117278 (117319.88 )& 27 \\
    sc-shipsec5 & 147140 (147179.12) & 147656 (147710.75) & 146991 (147022.95)& 149 \\ \hline
    soc-BlogCatalog & 20752 (20752) &20752 (20752.01)  & 20752 (20752.01)& 0 \\
    soc-brightkite & 21190 (21190) & 21190 (21190.09) & 21190 (21190.09)& 0 \\
    soc-buzznet & 30625 (30625) & 30613 (30613) & 30613 (30613)& 12 \\
    soc-delicious & 85660 (85696.77) & 85343 (85364.83) & 85319 (85333.75)& 341 \\
    soc-digg & 103243 (103244.72) & 103234 (103234.01)  & 103234 (103234.01)& 9 \\
    soc-epinions & 9757 (9757) & 9757 (9757.02) & 9757 (9757.02)& 0 \\
    soc-flickr & 153272 (153272.03) &153271 (153274.09)  & 153271 (153271.45)& 1  \\
    soc-flixster & 96317 (96317) & 96317 (96317.02) & 96317 (96317.02)& 0\\
    soc-FourSquare & 90108 (90109.09) & 90108 (90108.13) & 90108 (90108.13)& 0 \\
    soc-gowalla & 84222 (84222.36) & 84222 (84224.28) & 84222 (84222.07)& 0 \\
    soc-livejournal & 	1869044 (1869054.64) & 	1868997(1869010.13) & 1868924 (1868932.92)& 120 \\
    soc-pokec & 843419 (843432.58) & 843768 (843783.01) & 843344 (843347.38)& 75 \\
    soc-youtube & 146376 (146376.13) & 146376 (146376.35) & 146376 (146376.1)& 0 \\
    soc-youtube-snap & 276945 (276945) & 276945 (276945.21) & 276945 (276945.21)& 0 \\ \hline
    tech-as-skitter & 527161 (527204.59) & 525132 (525149.68) & 525086 (525099.14)& 2075 \\
    tech-RL-caida & 74924 (74940.83 ) & 74618 (74625.67) & 74607 (74615.25)& 317 \\\hline
    scc\_infect-dublin & 9104 (9104) & 9110 (9112.56 ) & 9103 (9103)& 1 \\
    scc\_retweet-crawl & 8419 (8419) & 8419 (8419.02) & 8419 (8419.02)& 0 \\ \hline
    web-arabic-2005 & 114425 (114427.28) & 114431 (114435.40) & 114420 (114420.67)& 5 \\
    web-BerkStan & 5384 (5384) &5388 (5388.13)  & 5384 (5384.13)& 0 \\
    web-it-2004 & 414671 (414675.12) & 414854 (414874.98) & 414646 (414649)& 25 \\
    web-spam & 2298 (2298.01) & 2297 (2298.07) & 2297 (2297.26)& 1 \\
    web-wikipedia2009 & 648315 (648321.83) & 648385 (648401.24) & 648300 (648312.39)& 15\\ \hline

    \hline
    \end{tabular}
    	
        \label{tab:2}
\end{table}

From the results in Tables \ref{tab:1} and \ref{tab:2}, we observe that:

1) $\lincom$ attains the best known solutions for most instances, and makes a significant progress. 
In Fact, among all the 136 tested instances $\lincom$ has found covers with 26 less vertices on average.
This improvement is big, since it rarely happens to find a better solution \cite{DBLP:conf/ijcai/Cai15}.

2) $\lincom$ is more robust. Actually out of 12 classes, $\lincom$ outperforms FastVC over 7 classes, while FastVC outperforms $\lincom$ over 1 class ($\eg$, facebook networks).
It seems that our local search is too greedy and not as effective as FastVC for facebook networks.

3)
There are quite a few instances ($\eg$, soc-delicious) where InitVC outperforms FastVC.
This illustrates that our $InitVC$ procedure generates desired starting vertex covers.

Furthermore, the solutions to the following 9 instances are guaranteed to be optimal: ca-CSphd, ca-Erdos992, ia-email-EU, ia-reality, ia-wiki-Talk, soc-douban, soc-LiveMocha, soc-twitter-follows, tech-internet-as. So our $InitVC$ procedure is sometimes complete in practice.

\section{Conclusions and Future Work}
In this paper, we have developed a local search algorithm for MinVC called $\lincom$, based on reduction rules and data structures.
The reduction rules help generate a better quality starting vertex cover, while the data structures lower the complexities of the heuristics.

The main contributions are two folds:
(1) we have lowered the complexity of two vertex cover construction heuristics and the best-picking heuristic based on the score-based alternative partitions at the theoretical level;
(2) we apply these results and some reduction rules to develop a local search solver which outperforms the state-of-the-art.

As for future works we will utilize various diversification strategies to in our solver.
Also, we will apply reduction rules to select vertices for exchanging in local search.

\newpage
\bibliographystyle{aaai}
\bibliography{aaai-mvc}
\end{document}